\documentclass[10pt,journal,compsoc]{IEEEtran}

\ifCLASSOPTIONcompsoc
  \usepackage[nocompress]{cite}
\else
  \usepackage{cite}
\fi

\usepackage{booktabs}
\ifCLASSINFOpdf
   \usepackage[pdftex]{graphicx}
\else
\fi
\graphicspath{{Figures/png/}} 

\usepackage{amsmath}
\usepackage{multirow}
\usepackage{pbox}

\begin{document}
%
\title{Hermite-Gaussian Mode Detection via Convolution Neural Networks}



%
%
%
%

        

 \author{\IEEEauthorblockN{L.R. Hofer\IEEEauthorrefmark{1},
L.W.~Jones\IEEEauthorrefmark{1}, 
J.L.~Goedert\IEEEauthorrefmark{1}, 
and R.V.~Dragone\IEEEauthorrefmark{1}}\\
\IEEEauthorblockA{\IEEEauthorrefmark{1}DataRay Inc., 1675 Market St., Redding, CA, 96001, USA}
\thanks{Corresponding author: L.R. Hofer (email: lhofer@dataray.com).
This work was supported by DataRay Inc.}}

%
%

\markboth{}%
{Shell \MakeLowercase{\textit{et al.}}: Bare Advanced Demo of IEEEtran.cls for IEEE Computer Society Journals}
\IEEEtitleabstractindextext{%
\begin{abstract}
Hermite-Gaussian (HG) laser modes are a complete set of solutions to the free-space paraxial wave equation in Cartesian coordinates and represent a close approximation to physically-realizable laser cavity modes. Additionally, HG modes can be mode-multiplexed to significantly increase the information capacity of optical communication systems due to their orthogonality. Since, both cavity tuning and optical communication applications benefit from a machine vision determination of HG modes, convolution neural networks were implemented to detect the lowest twenty-one unique HG modes with an accuracy greater than 99\%. As the effectiveness of a CNN is dependent on the diversity of its training data, extensive simulated and experimental datasets were created for training, validation and testing.

\end{abstract}

}

\maketitle

\IEEEdisplaynontitleabstractindextext

%
\IEEEpeerreviewmaketitle

\ifCLASSOPTIONcompsoc
\IEEEraisesectionheading{\section*{Introduction}\label{sec:introduction}}
\else
\label{sec:introduction}
\fi 
Convolution neural networks (CNN) \cite{lecun1998gradient} have seen a resurgence in the last decade \cite{Krizhevsky2012} due to their ability to classify images with near human or better than human accuracy \cite{he2015delving}. These developments have revolutionized machine vision applications from cancer detection \cite{cirecsan2013mitosis} to optics \cite{dosovitskiy2015flownet}. One area of optics research in which CNNs are proving useful \cite{lin2018application} is laser beam profiling—where either a CCD or CMOS camera is used to determine the centroid, radius \cite{hofer2017scale} and quality of a laser beam (M$^2$) among other metrics. Since, a laser beam's M$^2$ value is closely related to its modal content, determining the beam's dominant HG mode with a CNN is of significant interest. Furthermore, using CNNs to identify the HG \cite{kogelnik1966laser} mode of a beam has applications ranging from optical communications to atomic physics. 

Mode-multiplexing can significantly increase the information capacity of optical communication systems\cite{bozinovic2013terabit, wang2012terabit} through use of Hermite-Gaussian (HG) and Laguerre-Gaussian (LG) modes whose respective constituent modes propagate independently of one another due to their orthogonality---HG as well as LG modes comprise a complete orthogonal basis set \cite{lasers}. Much attention has been given to modes that carry orbital angular momentum (OAM); however, Zhao \textit{et al.} suggested that OAM modes do not necessarily increase the information capacity of a system in comparison to non-OAM modes and furthermore, OAM modes are often more adversely affected by turbulence\cite{zhao2015capacity}. Although Trichili \textit{et al.} showed that the full LG basis set of modes can encode information when multiplexing and demultiplexing data \cite{Trichili2016}, HG modes propagate information in a free-space optical communication network with an equal information capacity \cite{chen2016there} to LG modes and can experience lower mode loss and lower mode cross-talk \cite{Ndagano2017, cox2019resilience}. Since previous work has demonstrated the ability of deep neural networks to identify both OAM \cite{Krenn2014, Krenn2016, Doster2017, tian2018turbo} and LG modes \cite{Lohani2018}, extending the use of CNNs to identify HG modes provides another route to mode-multiplexing with potentially lower error rates. 

Even though higher-order HG modes are useful in optical communications, they are problematic in optical setups that require only the fundamental TEM$_{00}$ \cite{ross2013laser} mode. As an example, self-built external cavity diode lasers \cite{macadam1992narrow}, often found in atomic physics labs, rely on a laser diode and diffraction grating which form an external cavity. The laser must be carefully tuned via temperature, current and grating position to produce the correct frequency and TEM$_{00}$ spatial mode. Since the laser can mode-hop to oscillate in higher transverse (HG) modes \cite{sivaprakasam1996mode, saliba2009mode} during tuning, any automated tuning procedure would require a determination of the HG mode. This makes a machine vision HG mode characterization tool eminently useful. 

In this paper, convolution neural networks (CNN) are used to accurately determine the Hermite-Gaussian mode of a laser beam. As CNNs require substantial amounts of labeled data, two separate datasets were created. First, a simulated dataset was generated for both training and validation of the CNN, whereas a second experimental dataset was created---using a spatial light modulator (SLM) and beam profiler---to test the CNN's ability to generalize to new data and adapt to experimental conditions. The mathematical form of HG modes is first described, followed by the simulated dataset, the SLM optical setup and the experimental dataset. Finally, the CNNs are detailed along with the training methods used to achieve the best classification of the HG modes.

\begin{figure}
\centering 
\includegraphics[width=.5\textwidth]{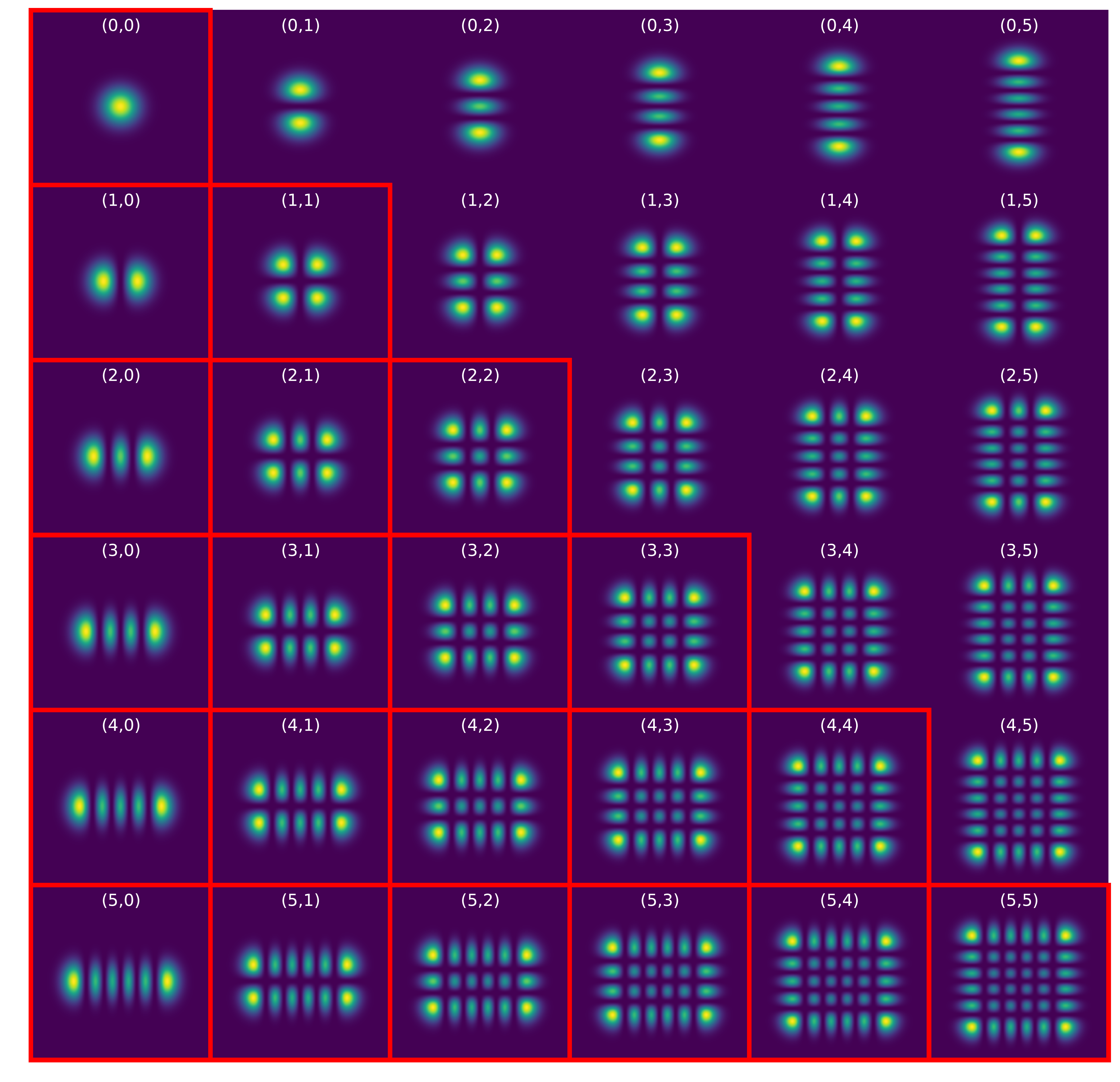} 
\caption[Table of Contents Figure Caption]{Intensity distributions of the lowest 36 Hermite-Gaussian modes. When considering rotational transformations, only 21 unique modes (bordered subplots) remain.} 
\label{fig:modes} 
\end{figure}

 In contrast to other state-of-the-art-mode detection techniques \cite{forbes2016creation}, which utilize computer generated holograms in all-optical setups \cite{schmidt2011real}, the method developed requires little to no optics and is additionally devoid of physically imposed constraints which limit the number of modes other methods can detect (e.g. the number of modes a single hologram can successfully demultiplex). Although single image evaluation times can be longer for the CNN mode detection method in comparison to all-optical techniques \cite{lyu2017fast}, recent advances in both CNN software architecture and processing chips designed for deep learning \cite{jouppi2017datacenter} should substantially lower evaluation times in the future.

\section*{Hermite-Gaussian Modes}\label{sec:hg}

The HG modes represent a set of solutions to the free-space paraxial wave equation in Cartesian coordinates \cite{lasers}. Along one dimension a HG mode's electric field is

\begin{align}
u_n(x, z)=&\left(\frac{2}{\pi}\right)^{\frac{1}{4}}\left( \frac{e^{-i(2n+1)\psi(z)}}{2^nn!w(z)}\right)^{\frac{1}{2}} \nonumber \\
&\times H_n\left(\frac{\sqrt{2}x}{w(z)}\right)e^{-i\frac{kx^2}{2R(z)}-\frac{x^2}{w^2(z)}}
\label{1DHE}
\end{align}

\noindent where $n$ is the mode of the higher-order beam, $k$ is the wave vector and $H_n$ is a Hermite polynomial of order $n$. The radius of the beam $w(z)$ at a given location $z$ along the axis of propagation is 

\begin{equation}
w(z)=w_0\left[1+\left(z/z_R\right)^2\right]^\frac{1}{2}
\end{equation}

\noindent where $w_0$ is the radius of the beam at the beam waist and the Rayleigh length is defined as $z_R=\pi w_0^2/\lambda$---with the wavelength of the beam denoted by $\lambda$. The Gouy phase is given by $\psi(z)=\tan^{-1}\left(z/z_R\right)$ and lastly, the radius of curvature is $R(z)=z\left[1+\left(z_R/z\right)^2\right]$. 

The HG mode's two-dimensional electric field is given by

\begin{equation}
u_{nm}(x, y, z)=u_n(x, z)u_m(y, z)
\label{eq:2DHE1}
\end{equation}

\noindent where $u_m(y, z)$ has a similar form to Eq.~\ref{1DHE}. The intensity distribution of the HG mode (see Fig.~\ref{fig:modes}) can be determined \cite{ross2013laser} via

\begin{equation}
I(x,y,z)=u_{nm}(x, y, z)u_{nm}^*(x, y, z)
\end{equation}

\noindent and the phase distribution $\phi(x, y, z)$ is calculated by taking the angle of $u_{nm}(x, y, z)$ in the complex plane.

\section*{Simulated Data}\label{sec:simulated}
Convolution neural networks require significant amounts of labeled data to properly train and thus a simulated dataset was generated using Eq.~\ref{eq:2DHE1}. A Python program was written to generate arbitrary HG mode electric field distributions from which their respective intensity and phase distributions were obtained. Since the accuracy of the CNN and its ability to generalize to new data increases with the diversity of its training set, the simulated data was generated to cover a parameter space consisting of the beam's radii along the major and minor axes ($w_{0x}$, $w_{0y}$), the beam's centroid ($x_0$, $y_0$) and the orientation of the beam $\theta$. The beam's amplitude was not included in the parameter space since the resulting image is normalized before passing into the CNN. Furthermore, the beams were simulated at the beam waist since the beam's position along the axis of propagation does not generate unique data. Rather than mapping the beam parameter space, each of the parameters was randomized within physically realizable bounds.

\begin{figure}
\centering 
\includegraphics[width=.5\textwidth]{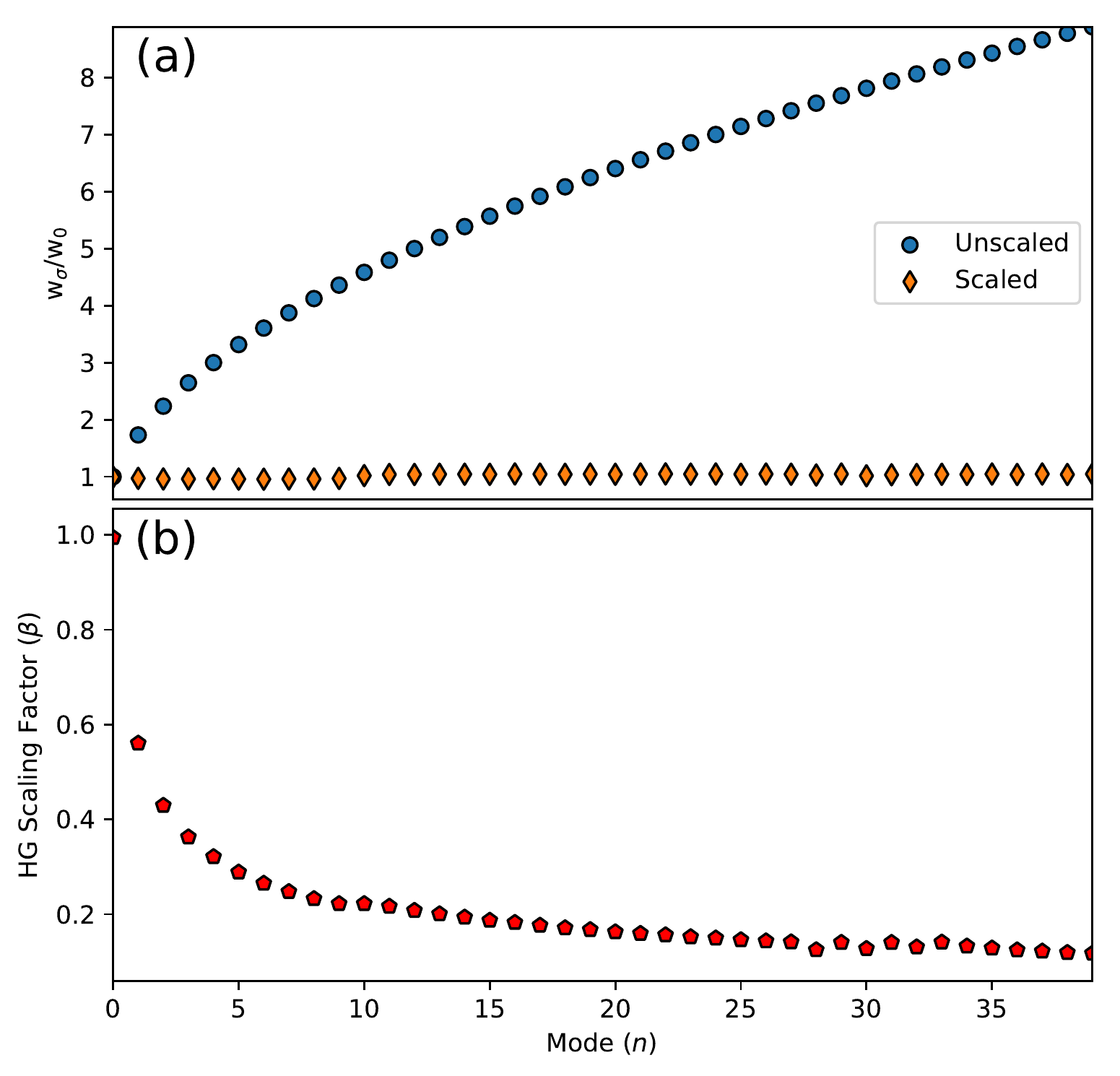} 
\caption[Table of Contents Figure Caption]{ (a) Ratio of the measured Hermite-Gaussian (HG) beam radius $w_{\sigma}$ and its respective input beam radius $w_0$ (see Eq.~\ref{1DHE}) as a function of the HG mode $n$ along a single axis. Both input radii with the HG scaling factor (diamonds) and without the scaling factor (circles) are shown. (b) The HG radius scaling factor $\beta$ as a function of the mode $n$.} 
\label{fig:scaling} 
\end{figure}

First, the bounds for the beam radii were determined.  To resolve an HG mode along one axis, a dark pixel should be seen on either side of each HG lobe thus requiring a minimum of $\left(2n+3\right)$ pixels---with the assumption that the lobe spacing is quasi-sinusoidal. This in turn results in a minimum input radius of

\begin{equation}
w_{0\text{min}}=\sqrt{2}p_w\left(2n+3\right)
\label{eq:wmin}
\end{equation}

\noindent where $\sqrt{2}p_w$ is the maximum distance (at a beam orientation of $\theta=\pi/4$) across a square pixel with width $p_w$. The maximum beam radius along both the major and minor axes is given by $w_{\text{max}}=s_l/3$, where $s_l$ is the simulated sensor size; larger radii would cause significant portions of the beam's power to be located off the simulated sensor. However, the HG beam radius increases with the mode order $n$ even though the input radius $w_0$ remains constant (see Fig.~\ref{fig:scaling}a) \cite{lasers}. Therefore, a scaling factor $\beta$ is numerically determined for each HG mode (see Fig.~\ref{fig:scaling}b) and multiplied with the desired output radius to give the correct input radius. Thus, the maximum input radius is

\begin{equation}
w_{0\text{max}}(n)=\frac{1}{3}s_l\beta(n)\text{.}
\end{equation}

\noindent Using $w_{0\text{min}}$ and $w_{0\text{max}}(n)$ a random radius can be generated for both the major and minor axes after which a random orientation for the beam is chosen with $0\leq \theta \leq 2\pi$.

After choosing the beam radii and orientation, valid bounds for the centroid are found such that the beam does not exceed the dimensions of the simulated sensor. Since the centroid values are given with respect to the image axes rather than the beam's major and minor axes, the beam radii along the image axes are calculated as follows

\begin{equation}
w_x=\pm\sqrt{w_{0x}^2\cos^2\theta+w_{0y}^2\sin^2\theta}
\end{equation}

\begin{equation}
w_y=\pm\sqrt{w_{0x}^2\sin^2\theta+w_{0y}^2\cos^2\theta}
\end{equation}

\noindent and the bounds for $x_0$ are then given by

\begin{equation}
x_{0\text{bounds}}=\pm\left(s_l/2-\alpha w_x\right)
\end{equation}

\noindent with a similar equation for the $y_0$ bounds except that $w_x$ is replaced by $w_y$. Due to the HG modes extending towards infinity, the radius is scaled by $\alpha=1.5$ such that a majority of the beam's power is incident on the simulated sensor. Using the centroid bounds, random values for the centroid are generated.

\begin{figure}
\centering 
\includegraphics[width=.45\textwidth]{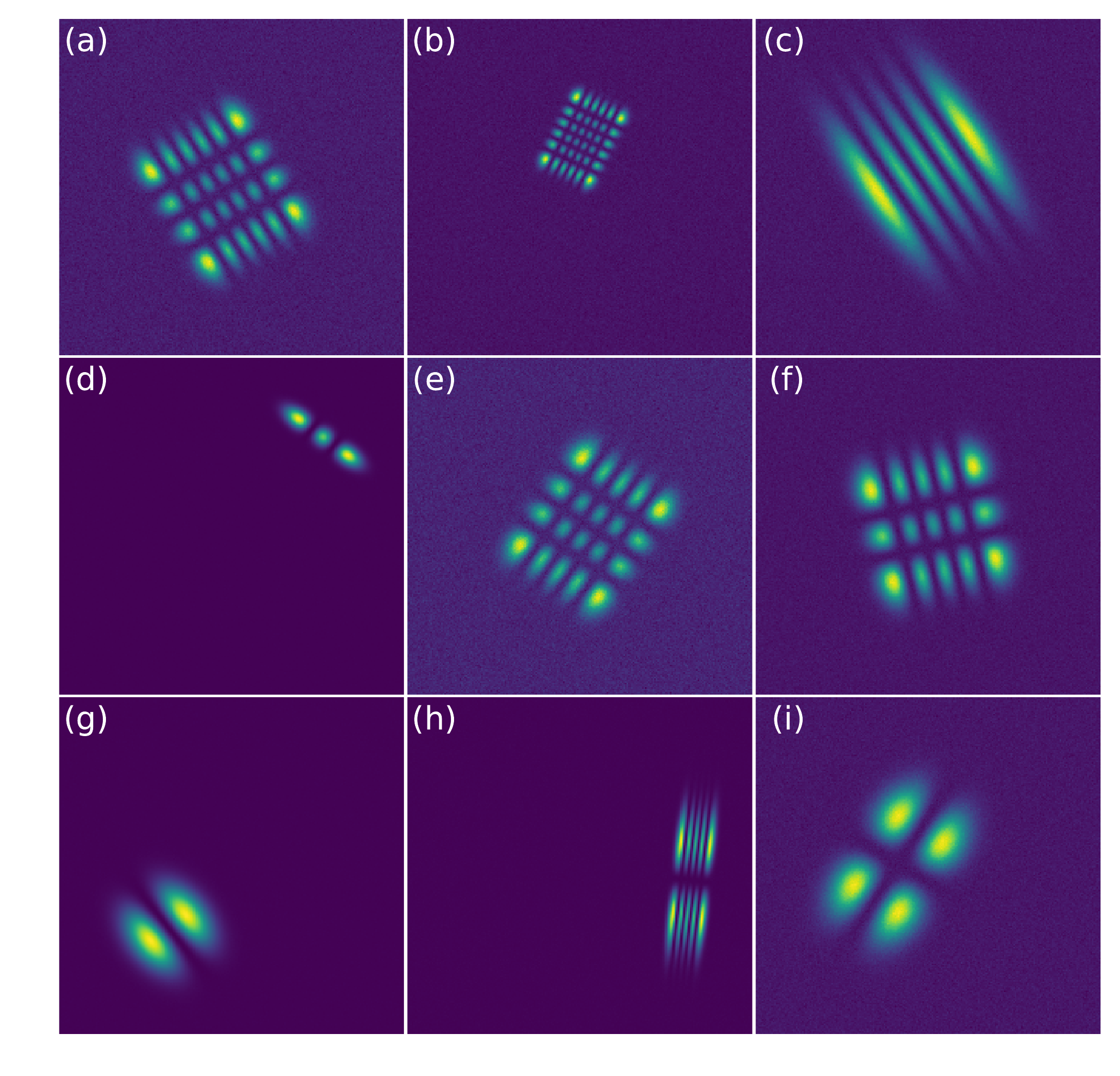} 
\caption[Table of Contents Figure Caption]{Subset of the simulated beams used to train and validate the convolution neural network. The simulated beams are given random values for the beam radii, beam centroid and orientation of the beam. Additionally, Gaussian noise is added to each image to approximate real-world conditions.}
\label{fig:simulated_data} 
\end{figure}

After generating a randomized beam, the maximum amplitude is scaled to one and Gaussian noise added to simulate experimental conditions. The standard deviation of the noise is itself randomly pulled from a Gaussian distribution which has a standard deviation of $\sigma=0.02$ and replicates real noise values seen on a sensor. After the noise has been added, the images are saved as PNGs (224$\times$224 pixels) which both compresses and scales the data between 0 and unity. A training dataset and a validation dataset (see Fig.~\ref{fig:simulated_data}) are generated with 300 and 200 images respectively for each of the lowest twenty-one unique HG modes (see Fig.~\ref{fig:modes}).

\section*{Experimental Data}
\begin{figure}
\centering 
\includegraphics[width=.5\textwidth]{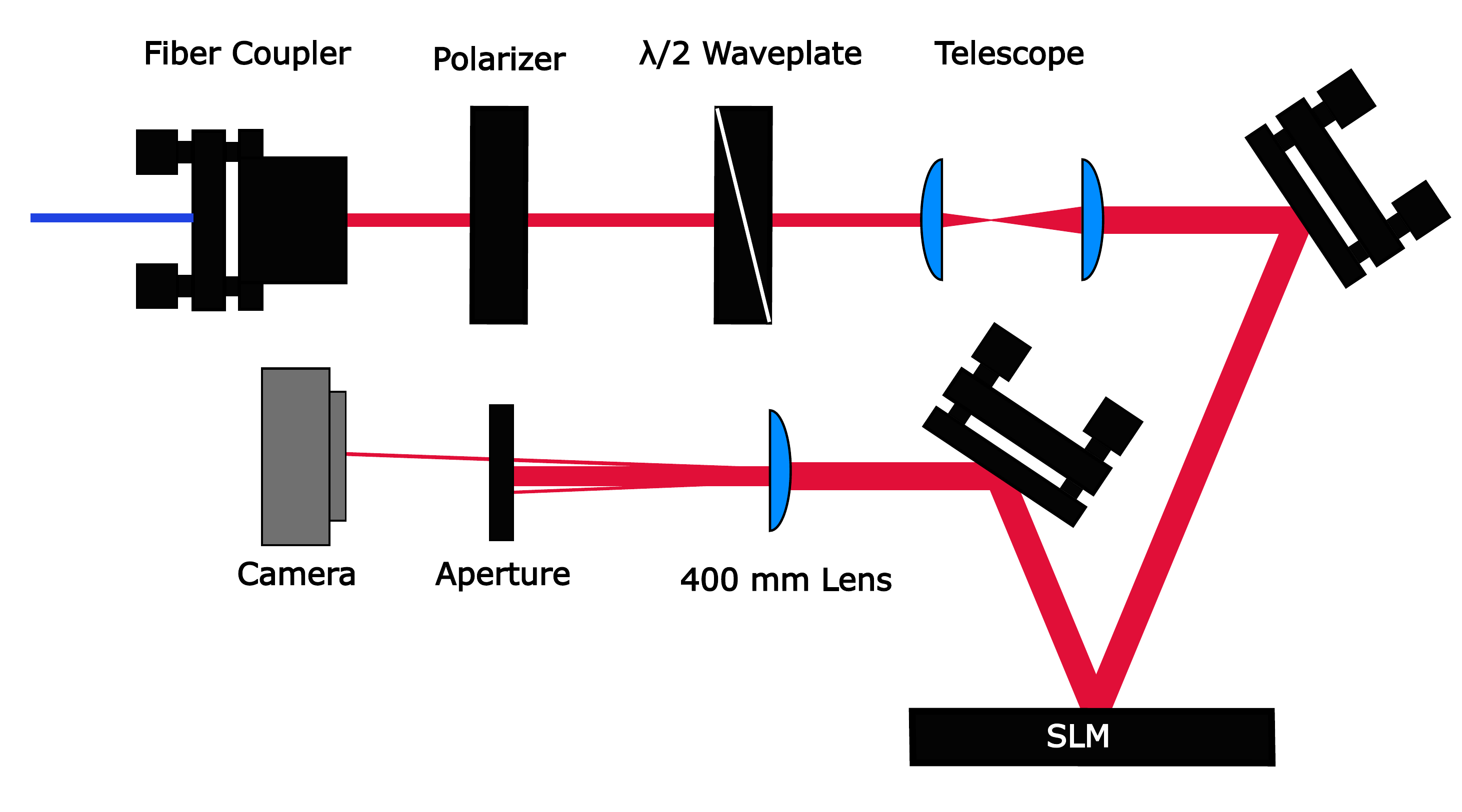} 
\caption[Table of Contents Figure Caption]{Optical setup for generating arbitrary Hermite-Gaussian beams. A 675 nm fiber-coupled laser beam exits the fiber-coupler and immediately passes through a polarizer followed by a $\lambda$/2 half-waveplate (HWP) both in rotation mounts. The beam is expanded through a series of lenses and is then incident on the SLM. The HWP preceding the SLM is rotated to align the beam polarization parallel with the vertical axis of the SLM. A complex-amplitude modulation hologram is applied to the SLM and the reflected beam then passes through a 400 mm lens which allows the various diffraction orders—due to the hologram's blazed grating—to be separated at the lens' focus. Finally, the first-order diffracted beam is separated via an aperture and imaged with a beam profiling camera.} 
\label{fig:experiment_setup} 
\end{figure}

To further validate the CNNs' effectiveness, an optical setup was constructed to create HG beams and acquire their images (see Fig.~\ref{fig:experiment_setup}). A single mode fiber-coupled laser with a 675 nm wavelength and an initial diameter of 1 mm was used as the source and passed first through a polarizer, which ensured the beam was linearly polarized along a single axis, followed by a lambda half-waveplate. The beam was then expanded to 9 mm in diameter and was incident on a spatial light modulator (SLM) at a slight angle with the preceding half-waveplate used to orient the beam's polarization parallel to the SLM's vertical axis. Computer generated holograms (CGH) with blazed gratings were loaded onto the SLM to create the HG modes. Following the SLM, the beam passed through a 400 mm aspheric focusing lens which separated the different diffraction orders---arising due to the CGH's blazed grating---near the focus and an aperture then allowed only the first-order beam to be imaged by the beam profiling camera.

\begin{figure}
\centering 
\includegraphics[width=.45\textwidth]{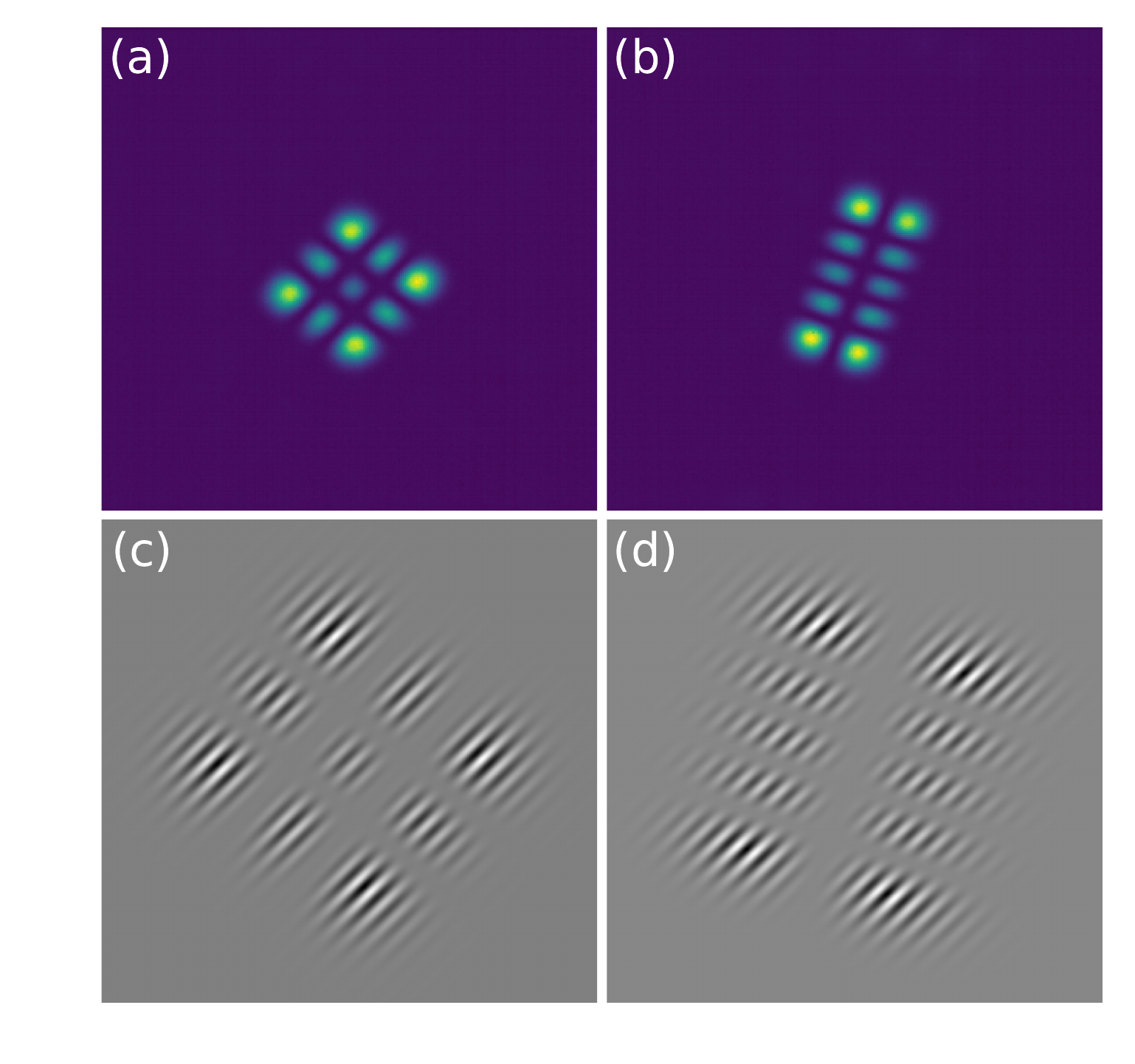} 
\caption[Table of Contents Figure Caption]{(a)-(b) Sample of the experimental dataset images used in testing the convolution neural networks with each image consisting of 256$\times$256 pixels. (c)-(d) The complex amplitude modulation (CAM) hologram used to generate the corresponding intensity distribution above. Note, the blazed grating frequency in each CAM hologram has been decreased by 80\% to increase its visibility.} 
\label{fig:exp_holo} 
\end{figure}

The optical setup centered on the spatial light modulator which enabled the phase of the electric field to be modulated on a per-pixel basis through the use of nematic liquid crystals \cite{konforti1988phase}. A Meadowlark Optics SLM was used with a resolution of 1920$\times$1152, square pixels of width 9.2 $\mu$m and a fill factor of 95.7\%. Although phase-only modulation holograms can be used to create higher-order modes \cite{matsumoto2008generation}, Arrizon \textit{et al.} demonstrated that both the phase and amplitude of the outgoing beam can be modulated with a phase-only SLM \cite{arrizon2007pixelated} through the use of complex amplitude modulation (CAM) holograms (see Fig.~\ref{fig:exp_holo}c-d). 

CAM holograms produce better quality HG modes than phase-only holograms \cite{200Modes} and were utilized to generate the experimental dataset (see Fig.~\ref{fig:exp_holo}a-b). The HG amplitude and phase distributions were calculated through Eq.~\ref{eq:2DHE1} and then used in the CAM hologram. Similar to the simulation dataset, the beam parameters of the holograms, including the beam radii and the orientation of the beam, were randomized to increase the diversity of the experimental dataset. Because the quality of the output HG beam deteriorated when the SLM's input beam was not centered on the hologram's centroid, the centroid position was kept constant.

The CAM holograms \cite{slmbook} contained a blazed grating which created various diffraction orders clearly seen at the focal plane of the lens following the SLM. The aperture near the focal plane isolated the first diffraction order---which contained the best representation of the HG beam---and an image was subsequently acquired by a beam profiling camera placed in the lens' Fourier-plane. During acquisition of the experimental dataset, the camera's software automatically adjusted the image exposure and between 115-120 randomized images with dimensions of 256$\times$256 pixels were recorded for each of the twenty-one HG modes. 

\section*{Convolution Neural Network}
In keeping with the simulation and data aquistion programs, a Pythonic approach to implementing the CNNs was taken with PyTorch's \cite{pytorch} deep learning framework. PyTorch has popular CNNs pretrained on the ImageNet dataset \cite{deng2009imagenet}, which allowed transfer learning \cite{yosinski2014transferable} to be taken advantage of and significantly shortened the training time required for a CNN. Although several successful CNN architectures have been put forth in recent years including AlexNet \cite{Krizhevsky2012}, VGG\cite{Simonyan2014} and ResNet \cite{He2016} among others \cite{huang2017densely, iandola2016squeezenet} ResNet was chosen, as it achieves more accurate results on the ImageNet dataset compared to older CNNs such as AlexNet. Additionally, ResNet has several million fewer parameters than a VGG CNN with comparable depth which decreases the training time and, when deployed, the evaluation time per image. However, ResNet itself has several implementations differing primarily with regards to the depth of the neural network and 18, 34, 50, 101, and 152 layer pretrained variants are available with PyTorch. Generally, a deeper neural network can achieve higher accuracies; however, deeper networks also have far more parameters and thus require longer training times, take more system memory, and increase single image evaluation times. Therefore, we opted to work with the smaller ResNet versions: ResNet18 and ResNet34. Furthermore, the CNNs were trained in the cloud with Google Compute Engine's Deep Learning virtual machine utilizing a Tesla P100 NVIDIA GPU. 

After choosing the ResNet models, several parameters were set to achieve maximum accuracy on the simulated and experimental data sets. Cross entropy loss was utilized as the loss function for every CNN trained, whereas both Adam \cite{kingma2014adam} (an adaptive optimizer) and stochastic gradient descent (SGD) were used for the optimizer function. However, after initial testing, we found, in keeping with recent literature \cite{wilson2017marginal}, that SGD was better able to generalize and achieved higher accuracy results on the experimental dataset than Adam. Thus, the following CNNs were trained using SGD with momentum as the optimizer. 

Lastly, data transforms were used throughout both the training and validation steps. Since increasing the diversity of the CNN's training data increases the CNN's ability to generalize to new data, two transforms were used during training. In the first transform, the images were randomly cropped between 0.08 and 1.0 of the initial image size after which they were given a random aspect ratio between 3/4 and 4/3 and finally resized to 224$\times$224 pixels. The second transform randomly flipped 50\% of the images along the horizontal axis before normalizing them and passing them into the CNN. During validation, the input data was not substantially altered; rather, the images were cropped to 224$\times$224 pixels about the center of the image to match the CNN's expected input size and then normalized before entering the CNN.

The image resolution of 224$\times$224 pixel was chosen to match the input layer of the pre-trained ResNet CNNs.
Although higher resolution images can increase a CNN's classification accuracy \cite{wu2015deep}, they simultaneously increase both training and single image evaluation times. Therefore, to mitigate potentially detrimental effects on the classification accuracy due to the image resolution, both the simulated and experimental datasets were constructed such that the dominant features (lobes) of the modes were always resolvable.

\section*{Results}

\begin{figure}
\centering 
\includegraphics[width=.5\textwidth]{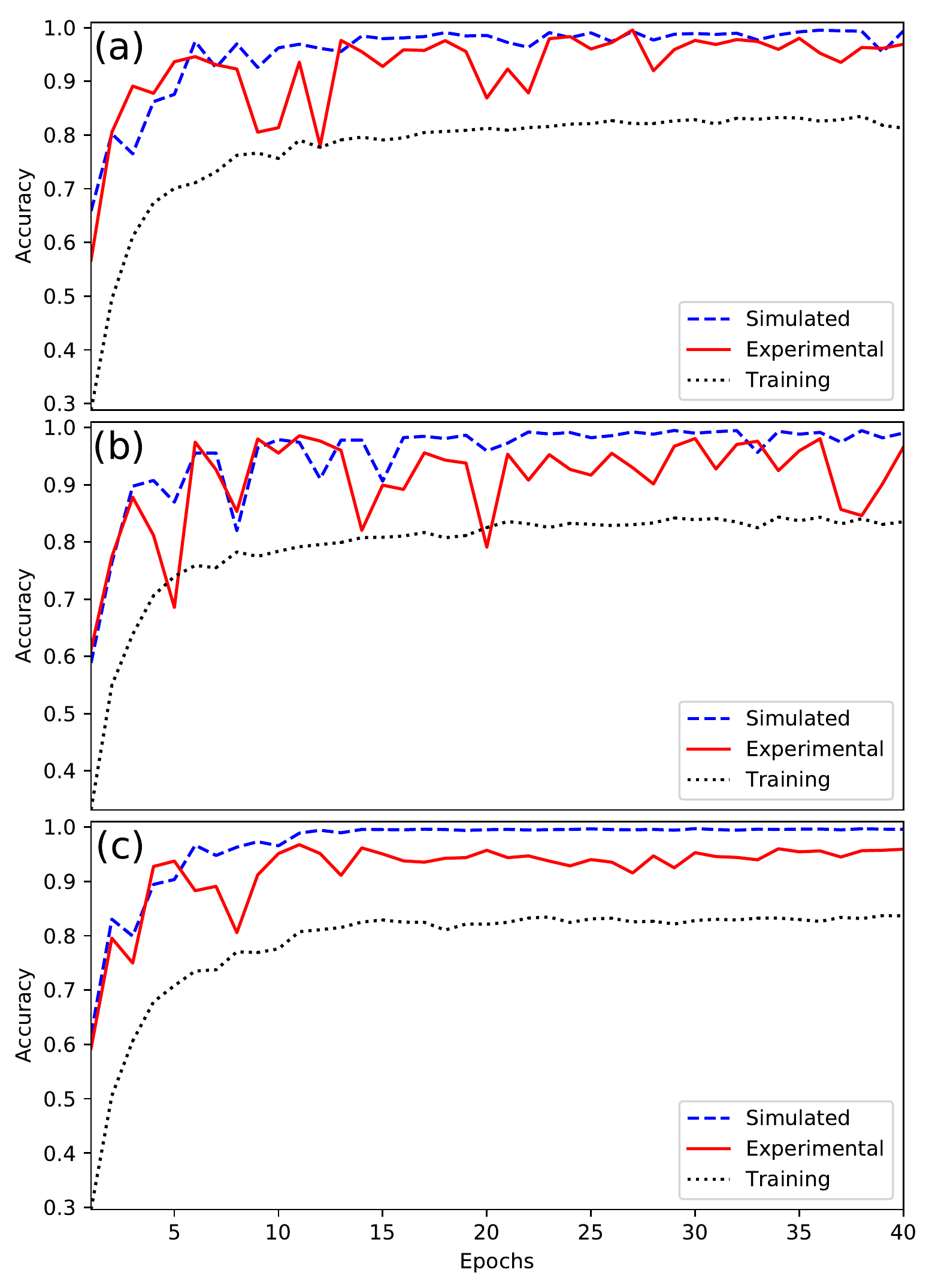} 
\caption[Table of Contents Figure Caption]{(a) Accuracy of a ResNet18 convolution neural network (CNN) as a function of the epoch with a constant learning rate of 0.001, momentum of $\mu$=0.9 and batch size of eight. The best accuracy achieved on the experimental dataset was 99.56\%. (b) Accuracy of a ResNet34 CNN as a function of the epoch for a constant learning rate of 0.0001. The momentum and batch size were the same as in (a), whereas the highest accuracy attained on the experimental dataset was 98.45\%. (c) Accuracy of the ResNet18 CNN as a function of the epoch with a batch size of eight and momentum of $\mu$=0.9. The learning rate was initially set to 0.001 and reduced by an order of magnitude every ten epochs. Although the maximum experimental accuracy (96.74\%) was reduced, it oscillated far less than (a) or (b) and was fairly asymptotic.} 
\label{fig:results} 
\end{figure}

Once the loss and optimizer functions along with the data transforms were determined, the CNNs hyperparameters including the batch size (in this case, the number of images fed into the CNN at a time), learning rate and momentum were tuned. Initially, the CNNs were trained for forty epochs (number of times the entire dataset is passed through the CNN) with a batch size of eight, a momentum of $\mu=0.9$ and constant learning rates of \{0.1, 0.01, 0.001, 0.0001\} (see Table~\ref{table:fixed_results}). A constant learning rate of 0.001 was optimal for ResNet18 and a maximal accuracy of 99.56\% was achieved on the experimental dataset, with a corresponding accuracy of 99.31\% on the simulation dataset (see Fig.~\ref{fig:results}a). For ResNet34 a constant learning rate of 0.0001 proved best (see Fig.~\ref{fig:results}b) yielding a maximum accuracy of 98.45\% on the experimental dataset and a coinciding accuracy of 99.29\% on the simulation dataset. ResNet18 alone was subsequently used as it achieved similar accuracy to ResNet34 and its smaller size decreased training and evaluation times.  After training, the ResNet18 evaluated images in approximately 100 milliseconds on a CPU and 5 milliseconds when utilizing the GPU.

\begin{table}
\centering
{\scriptsize
\begin{tabular}{lccccc}
\toprule
    Model &  Learning Rate &  Best Exp. Acc. (\%) &  Best Corr. Acc. (\%) &  Time (m) \\
\midrule
 ResNet18 &         0.1 &               31.19 &                66.10 &      43.1 \\
 ResNet18 &         0.01 &               56.46 &                91.07 &      39.8 \\
 ResNet18 &         0.001 &               99.56 &                99.31 &      39.5 \\
 ResNet18 &         0.0001 &               90.74 &                98.31 &      39.7 \\
 ResNet34 &         0.1 &               37.70 &                65.88 &      56.7 \\
 ResNet34 &         0.01 &               29.56 &                96.74 &      50.4 \\
 ResNet34 &         0.001 &               98.57 &                97.43 &      50.3 \\
 ResNet34 &         0.0001 &               98.45 &                99.29 &      50.3 \\
\bottomrule
\end{tabular}
}
  \caption{\label{table:fixed_results}The initial results of the convolution neural networks with fixed learning rates are shown. Each CNN has a batch size of eight and a momentum of $\mu$=0.9. A model's best accuracy on the experimental dataset along with the corresponding accuracy on the simulated dataset is given in addition to the total time (minutes) to train the model for forty epochs. Note that the highest accuracy ResNet18 and ResNet34 CNNs display oscillatory behavior around the asymptote (see Fig.~\ref{fig:results}a-b).}
\end{table}

Although the CNNs achieved fairly high results on both the simulated and experimental datasets, their accuracies on the experimental dataset failed to reach satisfactory asymptotes but rather oscillated substantially from epoch to epoch---indicating the hyperparameters were not properly tuned to target a local minimum. This is problematic if the training dataset is altered slightly. As an example, a second random simulated dataset was generated (utilizing the same bounds as the first), and used to train a set of CNNs with same hyperparameters as above. In this case, the best accuracy achieved for a ResNet18 CNN (learning rate of 0.001) dropped to 98.05\% and 99.31\% on the experimental and simulated datasets respectively.

To better target a local minimum during training, a step scheduler was employed in which the learning rate was decreased by an order of magnitude every ten epochs. This was tested for a series of CNNs with batch size eight, momentum of $\mu$=0.9 and initial learning rates of \{0.1, 0.01, 0.001, 0.0001\}. The best performing CNN (see Fig.~\ref{fig:results}c) had an initial learning rate of 0.001 and resulted in an accuracy of 96.74\% on the experimental dataset and 98.86\% on the simulation dataset. Even though the best ResNet18 CNN trained with a fixed learning rate had higher accuracies, the CNN trained with the step scheduler showed substantially lower amplitude oscillations around the asymptote demonstrating a local minimum was more effectively reached.

\begin{table}
\centering
{\scriptsize
\begin{tabular}{lccccc}
\toprule
 Learning Rate &  Momentum &  Batch Size &  Best Exp. Acc. (\%) &  Corr. Acc. (\%) \\
\midrule
      0.014367 &  0.864872 &   64 &               99.44 &           99.57 \\
      0.025743 &  0.357709 &   16 &               99.21 &           99.55 \\
      0.069024 &  0.135448 &   64 &               98.81 &           99.55 \\
      0.014746 &  0.283051 &   16 &               98.77 &           99.17 \\
      0.035026 &  0.776323 &   32 &               98.53 &           98.90 \\
\bottomrule
\end{tabular}
}
  \caption{\label{table:random_results}Results from the optimized ResNet18 convolution neural networks which utilized a hyperparameter random search in conjunction with a step scheduler. The initial learning rate was randomized between 0.1 and 0.001, the momentum was bounded by 0 and 1 and the batch size was set to 2$^l$, where $l$ was an integer given by $3\leq l \leq 8$. The best accuracy on the experimental dataset along with the corresponding accuracy on the simulation dataset is given for each model (see Fig.~\ref{fig:results2}a for accuracy vs. epoch).}
\end{table}

\begin{figure}
\centering 
\includegraphics[width=.5\textwidth]{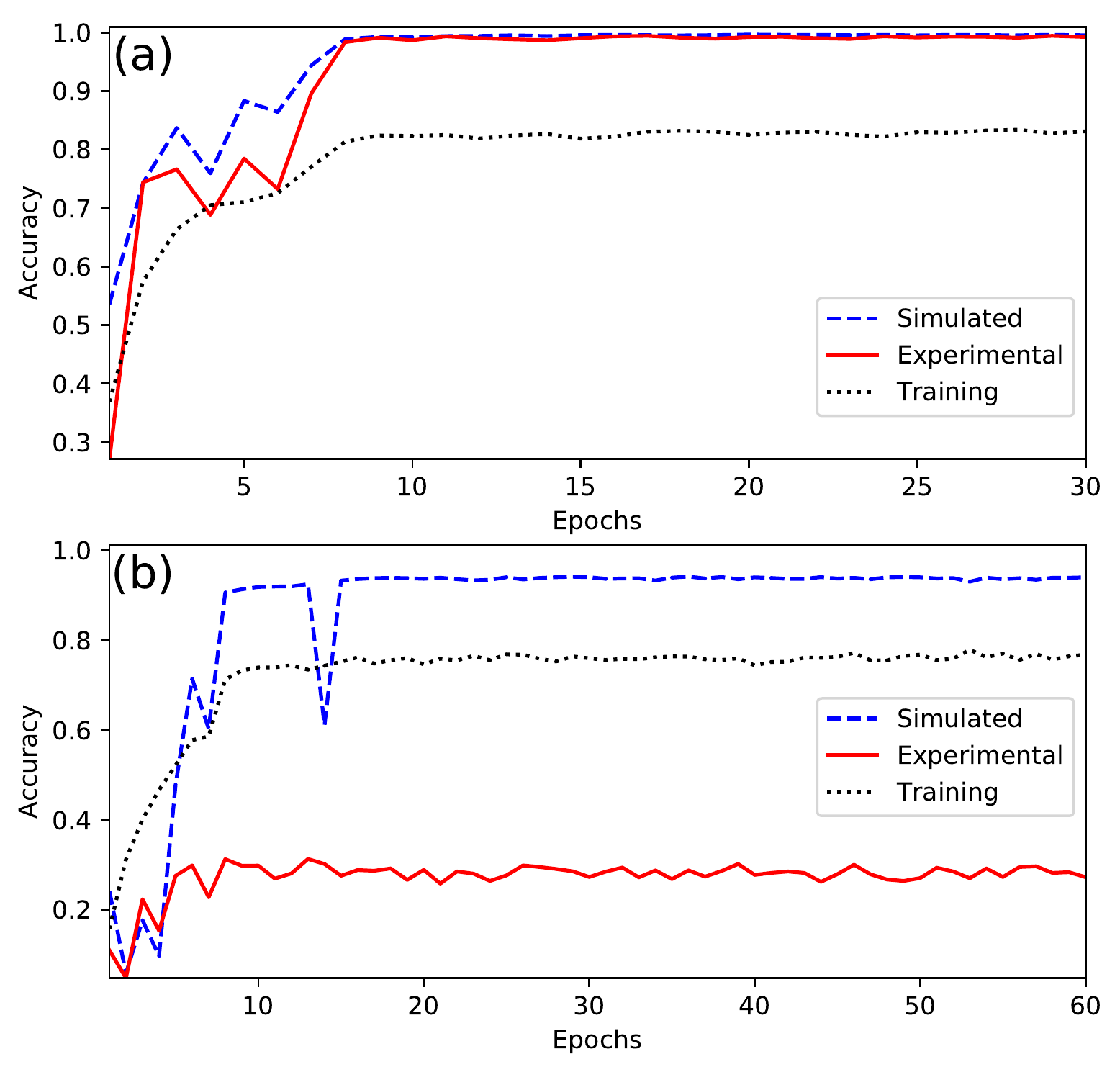} 
\caption[Table of Contents Figure Caption]{(a) Accuracy of a ResNet18 CNN as a function of the epoch with a batch size of 64 and momentum of $\mu$=0.864872. The learning rate was initially set to 0.0143673 and reduced by an order of magnitude every seven epochs. Note that the overall accuracy achieved---99.44\% on the experimental dataset, 99.57\% on the simulated dataset---was slightly better than the CNN trained with the fixed learning rate (see Fig.~\ref{fig:results}a) and furthermore displayed minimal oscillations around the asymptote indicating a local minimum was more effectively reached. (b) Accuracy of a ResNet18 CNN as a function of the epoch with the same hyperparameters as (a), but without pretraining on ImageNet. The best accuracy achieved was 31.31\% on the experimental dataset corresponding to 92.40\% on the simulation dataset; thus, pretraining in (a) significantly increased the CNN's accuracy in comparison to the non-pretrained CNN in (b).} 
\label{fig:results2} 
\end{figure}

In an effort to achieve both the high accuracy of the fixed learning rate run and the asymptotic behavior of the scheduled run, a random search \cite{bergstra2012random} of the hyperparameters was employed in conjunction with the step scheduler. The scheduler decreased the learning rate by a factor of ten every seven epochs and bounds were set for each hyperparameter with an initial learning rate between 0.1 and 0.001, a momentum of $0\leq \mu \leq 1$ and a batch size of $2^l$ where $l$ is an integer given by $3\leq l \leq 8$. Fifty different sets of hyperparameters were randomly chosen from within these bounds and then trained for thirty epochs each (see Table~\ref{table:random_results}). The CNN with the highest accuracy (learning rate 0.0143673, momentum $\mu$= 0.864872, batch size of 64) on the experimental data attained an accuracy of 99.44\% and a corresponding accuracy of 99.57\% on the simulated dataset (see Fig.~\ref{fig:results2}a). Thus, the overall accuracy was slightly higher than for the CNN trained with a fixed learning rate; however, just as importantly, the accuracy on both the simulation and experimental datasets become asymptotic indicating a local minimum was satisfactorily reached. 

Finally, the highest accuracy CNN from the random search was utilized without pretraining to determine the effect of transfer learning. The ResNet18 CNN was trained for sixty epochs with an initial learning rate of 0.0143673, momentum of $\mu$=0.864872 and batch size of 64 (see Fig.~\ref{fig:results2}b). The CNN obtained a maximum accuracy of 31.31\% on the experimental dataset and a corresponding accuracy of 92.40\% on the simulation dataset---which was substantially lower than pretrained model's accuracies for both datasets. Furthermore, the relative accuracy difference between the non-pretrained CNN and the pretrained CNN was significantly larger on the experimental dataset than the simulation dataset. This indicates that pretraining increased the overall accuracy of the CNN and additionally had an outsized impact on the CNN's ability to generalize to real world data.

\section*{Conclusion}
We have demonstrated that a convolution neural network (CNN) can be used to classify the lowest twenty-one unique Hermite-Gaussian (HG) modes with an accuracy of 99.44\%. The primary CNN used was an eighteen layer ResNet variant trained with a step scheduler for the learning rate and hyperparameters tuned with a random search. To facilitate in training the CNN, a large simulated dataset of HG modes was created in which each of the beam's parameters including orientation, centroid and radii were randomized within physically realizable bounds. Furthermore, an experimental dataset of HG modes was acquired through an optical setup utilizing a spatial light modulator and was used to test the CNN's ability to generalize to real-world data and experimental conditions. 

As stated previously, the trained CNN could be used to automatically tune the transverse HG mode output of a laser cavity or detect HG modes in optical communications. Since the current training dataset contains unique modes in random orientations, a beam under evaluation can be determined irrespective of orientation which is particularly useful in a laboratory setting. If the application of the CNN were solely optical communications, a dataset could also be constructed using all the modes (not only the unique ones); however, this would require a further bound on the orientation of the beams in the simulated dataset and additionally the beam would need to be correctly oriented with respect to the camera when the CNN is deployed.

In the future, a larger data set of modes could be both simulated and experimentally acquired. Moreover, superpositions of various HG modes could also be labeled as individual classes for input into the CNN. This would allow the data in a multiplexed beam to be determined without explicitly demultiplexing the beam with optics. However, this is bounded by the finite number of classes a CNN can accurately classify---which is problematic as the number of classes grows exponentially with the modes in the multiplexed beam.

\section*{Funding}
This work was supported by DataRay Inc.

\section*{Acknowledgments}
L. Hofer would like to thank both Alex Christoph and John Cadigan for helpful discussions on convolution neural networks.

\section*{References}

\bibliography{bibliography}{}

\begin{thebibliography}{10}

\bibitem{lecun1998gradient}
Yann LeCun, L{\'e}on Bottou, Yoshua Bengio, and Patrick Haffner.
\newblock Gradient-based learning applied to document recognition.
\newblock {\em Proceedings of the IEEE}, 86(11):2278--2324, 1998.

\bibitem{Krizhevsky2012}
Alex Krizhevsky, Ilya Sutskever, and Geoffrey~E Hinton.
\newblock Imagenet classification with deep convolutional neural networks.
\newblock In {\em Advances in Neural Information Processing Systems}, pages
  1097--1105, 2012.

\bibitem{he2015delving}
Kaiming He, Xiangyu Zhang, Shaoqing Ren, and Jian Sun.
\newblock Delving deep into rectifiers: Surpassing human-level performance on
  imagenet classification.
\newblock In {\em Proceedings of the IEEE International Conference on Computer
  Vision}, pages 1026--1034, 2015.

\bibitem{cirecsan2013mitosis}
Dan~C Cire{\c{s}}an, Alessandro Giusti, Luca~M Gambardella, and J{\"u}rgen
  Schmidhuber.
\newblock Mitosis detection in breast cancer histology images with deep neural
  networks.
\newblock In {\em International Conference on Medical Image Computing and
  Computer-assisted Intervention}, pages 411--418. Springer, 2013.

\bibitem{dosovitskiy2015flownet}
Alexey Dosovitskiy, Philipp Fischer, Eddy Ilg, Philip Hausser, Caner Hazirbas,
  Vladimir Golkov, Patrick Van Der~Smagt, Daniel Cremers, and Thomas Brox.
\newblock Flownet: Learning optical flow with convolutional networks.
\newblock In {\em Proceedings of the IEEE International Conference on Computer
  Vision}, pages 2758--2766, 2015.

\bibitem{lin2018application}
Chern-Sheng Lin, Yu-Chia Huang, Shih-Hua Chen, Yu-Liang Hsu, and Yu-Chen Lin.
\newblock The application of deep learning and image processing technology in
  laser positioning.
\newblock {\em Applied Sciences}, 8(9):1542, 2018.

\bibitem{hofer2017scale}
Lucas~R Hofer, Rocco~V Dragone, and Andrew~D MacGregor.
\newblock Scale factor correction for gaussian beam truncation in second moment
  beam radius measurements.
\newblock {\em Optical Engineering}, 56(4):043110, 2017.

\bibitem{kogelnik1966laser}
Herwig Kogelnik and Tingye Li.
\newblock Laser beams and resonators.
\newblock {\em Applied Optics}, 5(10):1550--1567, 1966.

\bibitem{bozinovic2013terabit}
Nenad Bozinovic, Yang Yue, Yongxiong Ren, Moshe Tur, Poul Kristensen, Hao
  Huang, Alan~E Willner, and Siddharth Ramachandran.
\newblock Terabit-scale orbital angular momentum mode division multiplexing in
  fibers.
\newblock {\em Science}, 340(6140):1545--1548, 2013.

\bibitem{wang2012terabit}
Jian Wang, Jeng-Yuan Yang, Irfan~M Fazal, Nisar Ahmed, Yan Yan, Hao Huang,
  Yongxiong Ren, Yang Yue, Samuel Dolinar, Moshe Tur, and A.
\newblock Terabit free-space data transmission employing orbital angular
  momentum multiplexing.
\newblock {\em Nature Photonics}, 6(7):488, 2012.

\bibitem{lasers}
Anthony~E. Siegman.
\newblock {\em Lasers}.
\newblock University Science Books, 1986.

\bibitem{zhao2015capacity}
Ningbo Zhao, Xiaoying Li, Guifang Li, and Joseph~M Kahn.
\newblock Capacity limits of spatially multiplexed free-space communication.
\newblock {\em Nature Photonics}, 9(12):822, 2015.

\bibitem{Trichili2016}
Abderrahmen Trichili, Carmelo Rosales-Guzm{\'a}n, Angela Dudley, Bienvenu
  Ndagano, Amine~Ben Salem, Mourad Zghal, and Andrew Forbes.
\newblock Optical communication beyond orbital angular momentum.
\newblock {\em Scientific Reports}, 6:27674, 2016.

\bibitem{chen2016there}
Mingzhou Chen, Kishan Dholakia, and Michael Mazilu.
\newblock Is there an optimal basis to maximise optical information transfer?
\newblock {\em Scientific Reports}, 6:22821, 2016.

\bibitem{Ndagano2017}
Bienvenu Ndagano, Nokwazi Mphuthi, Giovanni Milione, and Andrew Forbes.
\newblock Comparing mode-crosstalk and mode-dependent loss of laterally
  displaced orbital angular momentum and hermite--gaussian modes for free-space
  optical communication.
\newblock {\em Optics Letters}, 42(20):4175--4178, 2017.

\bibitem{cox2019resilience}
Mitchell~A Cox, Luthando Maqondo, Ravin Kara, Giovanni Milione, Ling Cheng, and
  Andrew Forbes.
\newblock The resilience of hermite-and laguerre-gaussian modes in turbulence.
\newblock {\em arXiv preprint arXiv:1901.07203}, 2019.

\bibitem{Krenn2014}
Mario Krenn, Robert Fickler, Matthias Fink, Johannes Handsteiner, Mehul Malik,
  Thomas Scheidl, Rupert Ursin, and Anton Zeilinger.
\newblock Communication with spatially modulated light through turbulent air
  across vienna.
\newblock {\em New Journal of Physics}, 16(11):113028, 2014.

\bibitem{Krenn2016}
Mario Krenn, Johannes Handsteiner, Matthias Fink, Robert Fickler, Rupert Ursin,
  Mehul Malik, and Anton Zeilinger.
\newblock Twisted light transmission over 143 km.
\newblock {\em Proceedings of the National Academy of Sciences},
  113(48):13648--13653, 2016.

\bibitem{Doster2017}
Timothy Doster and Abbie~T Watnik.
\newblock Machine learning approach to oam beam demultiplexing via
  convolutional neural networks.
\newblock {\em Applied Optics}, 56(12):3386--3396, 2017.

\bibitem{tian2018turbo}
Qinghua Tian, Zhe Li, Kang Hu, Lei Zhu, Xiaolong Pan, Qi~Zhang, Yongjun Wang,
  Feng Tian, Xiaoli Yin, and Xiangjun Xin.
\newblock Turbo-coded 16-ary oam shift keying fso communication system
  combining the cnn-based adaptive demodulator.
\newblock {\em Optics Express}, 26(21):27849--27864, 2018.

\bibitem{Lohani2018}
Sanjaya Lohani, Erin~M Knutson, Matthew O’Donnell, Sean~D Huver, and Ryan~T
  Glasser.
\newblock On the use of deep neural networks in optical communications.
\newblock {\em Applied Optics}, 57(15):4180--4190, 2018.

\bibitem{ross2013laser}
T~Sean Ross and Society of~Photo-optical Instrumentation~Engineers.
\newblock {\em Laser Beam Quality Metrics}.
\newblock SPIE Press Bellingham, 2013.

\bibitem{macadam1992narrow}
KB~MacAdam, A~Steinbach, and Carl Wieman.
\newblock A narrow-band tunable diode laser system with grating feedback, and a
  saturated absorption spectrometer for cs and rb.
\newblock {\em American Journal of Physics}, 60(12):1098--1111, 1992.

\bibitem{sivaprakasam1996mode}
S~Sivaprakasam, Ranita Saha, P~Anantha Lakshmi, and Ranjit Singh.
\newblock Mode hopping in external-cavity diode lasers.
\newblock {\em Optics Letters}, 21(6):411--413, 1996.

\bibitem{saliba2009mode}
Sebastian~D Saliba, Mark Junker, Lincoln~D Turner, and Robert~E Scholten.
\newblock Mode stability of external cavity diode lasers.
\newblock {\em Applied Optics}, 48(35):6692--6700, 2009.

\bibitem{forbes2016creation}
Andrew Forbes, Angela Dudley, and Melanie McLaren.
\newblock Creation and detection of optical modes with spatial light
  modulators.
\newblock {\em Advances in Optics and Photonics}, 8(2):200--227, 2016.

\bibitem{schmidt2011real}
Oliver~A Schmidt, Christian Schulze, Daniel Flamm, Robert Br{\"u}ning, Thomas
  Kaiser, Siegmund Schr{\"o}ter, and Michael Duparr{\'e}.
\newblock Real-time determination of laser beam quality by modal decomposition.
\newblock {\em Optics Express}, 19(7):6741--6748, 2011.

\bibitem{lyu2017fast}
Meng Lyu, Zhiquan Lin, Guowei Li, and Guohai Situ.
\newblock Fast modal decomposition for optical fibers using digital holography.
\newblock {\em Scientific Reports}, 7(1):6556, 2017.

\bibitem{jouppi2017datacenter}
Norman~P Jouppi, Cliff Young, Nishant Patil, David Patterson, Gaurav Agrawal,
  Raminder Bajwa, Sarah Bates, Suresh Bhatia, Nan Boden, Al~Borchers, et~al.
\newblock In-datacenter performance analysis of a tensor processing unit.
\newblock In {\em 2017 ACM/IEEE 44th Annual International Symposium on Computer
  Architecture (ISCA)}, pages 1--12. IEEE, 2017.

\bibitem{konforti1988phase}
Naim Konforti, Emanuel Marom, and S-T Wu.
\newblock Phase-only modulation with twisted nematic liquid-crystal spatial
  light modulators.
\newblock {\em Optics letters}, 13(3):251--253, 1988.

\bibitem{matsumoto2008generation}
Naoya Matsumoto, Taro Ando, Takashi Inoue, Yoshiyuki Ohtake, Norihiro Fukuchi,
  and Tsutomu Hara.
\newblock Generation of high-quality higher-order laguerre-gaussian beams using
  liquid-crystal-on-silicon spatial light modulators.
\newblock {\em JOSA A}, 25(7):1642--1651, 2008.

\bibitem{arrizon2007pixelated}
Victor Arriz{\'o}n, Ulises Ruiz, Rosibel Carrada, and Luis~A Gonz{\'a}lez.
\newblock Pixelated phase computer holograms for the accurate encoding of
  scalar complex fields.
\newblock {\em JOSA A}, 24(11):3500--3507, 2007.

\bibitem{200Modes}
Carmelo Rosales-Guzm{\'a}n, Nkosiphile Bhebhe, Nyiku Mahonisi, and Andrew
  Forbes.
\newblock Multiplexing 200 spatial modes with a single hologram.
\newblock {\em Journal of Optics}, 19(11):113501, 2017.

\bibitem{slmbook}
Carmelo Rosales-Guzm{\'a}n and Andrew Forbes.
\newblock {\em How to shape light with spatial light modulators}.
\newblock SPIE Press, 2017.

\bibitem{pytorch}
Adam Paszke, Sam Gross, Soumith Chintala, Gregory Chanan, Edward Yang, Zachary
  DeVito, Zeming Lin, Alban Desmaison, Luca Antiga, and Adam Lerer.
\newblock Automatic differentiation in pytorch.
\newblock {\em OpenReview}, 2017.

\bibitem{deng2009imagenet}
Jia Deng, Wei Dong, Richard Socher, Li-Jia Li, Kai Li, and Li~Fei-Fei.
\newblock Imagenet: A large-scale hierarchical image database.
\newblock In {\em Computer Vision and Pattern Recognition, 2009. CVPR 2009.
  IEEE Conference on}, pages 248--255. Ieee, 2009.

\bibitem{yosinski2014transferable}
Jason Yosinski, Jeff Clune, Yoshua Bengio, and Hod Lipson.
\newblock How transferable are features in deep neural networks?
\newblock In {\em Advances in Neural Information Processing Systems}, pages
  3320--3328, 2014.

\bibitem{Simonyan2014}
Karen Simonyan and Andrew Zisserman.
\newblock Very deep convolutional networks for large-scale image recognition.
\newblock {\em arXiv preprint arXiv:1409.1556}, 2014.

\bibitem{He2016}
Kaiming He, Xiangyu Zhang, Shaoqing Ren, and Jian Sun.
\newblock Deep residual learning for image recognition.
\newblock In {\em Proceedings of the IEEE Conference on Computer Vision and
  Pattern Recognition}, pages 770--778, 2016.

\bibitem{huang2017densely}
Gao Huang, Zhuang Liu, Laurens Van Der~Maaten, and Kilian~Q Weinberger.
\newblock Densely connected convolutional networks.
\newblock In {\em CVPR}, volume~1, page~3, 2017.

\bibitem{iandola2016squeezenet}
Forrest~N Iandola, Song Han, Matthew~W Moskewicz, Khalid Ashraf, William~J
  Dally, and Kurt Keutzer.
\newblock Squeezenet: Alexnet-level accuracy with 50x fewer parameters and< 0.5
  mb model size.
\newblock {\em arXiv preprint arXiv:1602.07360}, 2016.

\bibitem{kingma2014adam}
Diederik~P Kingma and Jimmy Ba.
\newblock Adam: A method for stochastic optimization.
\newblock {\em arXiv preprint arXiv:1412.6980}, 2014.

\bibitem{wilson2017marginal}
Ashia~C Wilson, Rebecca Roelofs, Mitchell Stern, Nati Srebro, and Benjamin
  Recht.
\newblock The marginal value of adaptive gradient methods in machine learning.
\newblock In {\em Advances in Neural Information Processing Systems}, pages
  4148--4158, 2017.

\bibitem{wu2015deep}
Ren Wu, Shengen Yan, Yi~Shan, Qingqing Dang, and Gang Sun.
\newblock Deep image: Scaling up image recognition.
\newblock {\em arXiv preprint arXiv:1501.02876}, 2015.

\bibitem{bergstra2012random}
James Bergstra and Yoshua Bengio.
\newblock Random search for hyper-parameter optimization.
\newblock {\em Journal of Machine Learning Research}, 13(Feb):281--305, 2012.

\end{thebibliography}
\bibliographystyle{unsrt}

\end{document}